\documentclass[10pt]{iopart}

\usepackage{graphicx}
\usepackage{iopams}  
\usepackage{array}
\usepackage{url}
\newcolumntype{L}{>{$}l<{$}} 
\begin{document}

\title{Direct coupling of first-principles calculations with replica exchange Monte Carlo sampling of ion disorder in solids}

\author{Shusuke Kasamatsu}
\author{Osamu Sugino}

\address{The Institute for Solid State Physics, the University of Tokyo, 5-1-5 Kashiwanoha, Kashiwa-shi, Chiba 277-8581, Japan}
\ead{kasamatsu@issp.u-tokyo.ac.jp}
\vspace{10pt}
\begin{indented}
\item[] September 2018
\end{indented}

\begin{abstract}
We demonstrate the feasibility of performing sufficient configurational sampling of disordered oxides directly from first principles without resorting to the use of fitted models such as cluster expansion. This is achieved by harnessing the power of modern-day cluster supercomputers using the replica exchange Monte Carlo method coupled directly with structural relaxation and energy calculation performed by density functional codes. The idea is applied successfully to the calculation of the temperature-dependence of the degree of inversion in the cation sublattice of MgAl$_2$O$_4$ spinel oxide.
The possibility of bypassing fitting models will lead to investigation of disordered systems where cluster expansion is known to perform badly: for example, systems with large lattice deformation due to defects, or systems where long-range interactions dominate such as electrochemical interfaces.
\end{abstract}

%
%
%
%
\ioptwocol

\section{Introduction}
Functional materials including oxides and alloys have varying degrees of disorder that impact the physical properties and thus the performance of materials when applied to energy conversion, catalysis, thermal barrier coatings, structural materials, etc. Although first-principles calculations based on density functional theory (DFT) have proved to be quite a powerful tool for investigating such materials \cite{Hafner2006,Ceder2010}, single-point DFT calculations are insufficient for predicting disorder and its effect on various physical properties. For that, one needs to perform sufficient thermodynamical sampling based on statistical mechanics. Sampling with first-principles molecular dynamics (MD) is limited by the realistic time scale of the relaxation dynamics, so it is nearly useless for evaluating ion disorder in solids with rather high energy barriers between different configurations. On the other hand, Monte Carlo (MC) approaches can utilize `unphysical' trial steps such as the swapping of atom positions and are inherently free from such a limitation. However, MC simulations can still get trapped in local minima, and in practice, it is usually not computationally feasible to perform sufficient MC sampling in solid state systems directly based on DFT energies. This is reflected in the fact that  reports of such attempts are limited to relatively simple and quickly-relaxing systems that can be sampled just as well using first-principles MD (e.g., liquid Li or small molecular clusters \cite{Jellinek1998, Wang2004b}); we could find none that targeted solid-state alloy systems. To deal with this situation,  many workers have resorted to DFT-fitted models such as cluster expansion \cite{Sanchez1984,Sanchez2017, Wu2016} or lattice gas models \cite{DeGironcoli1991} for providing enough sampling points in MC simulations of functional materials with ion disorder. We also note that in a slightly different context of rare-gas melting physics, up to three-body potentials have been fitted to CCSD(T) level of theory to perform MC simulations \cite{Pahl2008}. Such fitting approaches, however, face limitations as the complexities increase in the material under study, as accurate mapping of DFT results to models with low calculation cost can be quite challenging for systems with complex long-range (e.g., Coulombic or dispersion) interactions \cite{Seko2014}. Lattice relaxation effects, which are rather pronounced in oxides with defects, are also known to decrease the accuracy of derived cluster expansion Hamiltonians \cite{Nguyen2017}. And such interactions are especially important for multivalent ion oxides and various heterointerfaces that are central to many applications of functional materials. 

We note that when a moderately accurate but inexpensive potential is available for the system under study, MC simulation with such a potential can be ``nested" between MC steps taken with DFT energies to speed up the sampling. The efficiency of this ``nested Monte Carlo (NMC)'' method \cite{Leiding2014,Leiding2016} depends on the quality of the potential, and thus the method does not allow the user to fully escape from the difficulties mentioned above. 

In view of such difficulties, we decided to reexamine the feasibility of bypassing fitting models and performing direct thermodynamical sampling  on DFT energies. While such an undertaking may have been viewed as out of the question a decade ago in all but the simplest systems, the steady increase in computational power available to researchers have changed the situation. Algorithms in computational statistical mechanics have evolved to take advantage of massively parallel supercomputers now available such as the replica exchange (RX) \cite{Swendsen1986, Hukushima1996, Sugita1999, Earl2005} and population annealing \cite{Hukushima2003,Machta2011} methods, as well as parallel multicanonical \cite{Zierenberg2013,Gross2017} or Wang-Landau methods \cite{Vogel2013}. Machine learning techniques and `materials informatics' have also emerged as possible game-changers due to the ability to generate huge amounts of data. 

In this work, we present an implementation of a Python framework for replica exchange Monte Carlo (RXMC) sampling coupled directly with local structural relaxation and energy calculation using DFT codes. The RXMC technique was first applied to spin glasses \cite{Hukushima1996}, and later combined with classical molecular dynamics to examine protein folding \cite{Sugita1999}, although we could not find any literature on its application to realistic solid state systems. Here, we benchmark the feasibility of this scheme on the first-principles calculation of the temperature-dependence of the degree of inversion in the cation sublattice of MgAl$_2$O$_4$ spinel oxide. We note that the importance of long-range interactions have been stressed in this material; overfitting of small-unit cell DFT results with a cluster expansion Hamiltonian resulted in a spurious prediction of a discontinuous order-disorder transition \cite{Seko2014}. Although careful selection of long-period input DFT structures was shown to alleviate this problem, such adjustment techniques require additional expertise. Moreover, if it is necessary to perform DFT calculations on long-period structures, the merit of reduced computational cost by using fitted models becomes smaller. The direct approach described in the following is much simpler, although it is limited by the size of the supercell that can be calculated using DFT within a reasonable amount of time.
 
\section{Methodology}
In the following, we begin with a short review of the basics of the RXMC methodology and how we adapt it to canonical ensemble sampling of crystalline solids with site disorder. Then we will follow with the description of our parallel implementation for coupling RXMC with DFT.
\subsection{Metropolis Monte Carlo sampling}
For atomistic systems, thermodynamical sampling is usually performed by molecular dynamics (MD) or MC methods, or a hybrid of both. The former retains some dynamical information while the latter does not. On the other hand, MD is limited by realistic relaxation time scales that hinder the global exploration of the phase space, while MC is free from such problem due to the freedom in the choice of trial steps. Since solid state systems that we are interested in here are characterized by rather deep potential minima separated by large potential barriers, we choose the MC method over MD in this work. We also note that although we focus on canonical ensemble sampling in this work, it is straightforward to apply MC methods to the grand canonical or multicanonical ensemble methods, while that is not the case for MD. 

The usual Metropolis Monte Carlo algorithm \cite{Metropolis1953} proceeds as follows:
\begin{enumerate}
\item
Start with an initial crystal structure $\mathbf{X}$ with energy $E(\mathbf{X})$.
\item
Generate a trial structure $\mathbf{X}^\prime$ by perturbing a part of the previous structure and calculate the energy $E(\mathbf{X}^\prime)$.
\item 
Accept the trial structure as the next structure (set $\mathbf{X} \leftarrow \mathbf{X}^\prime$) according to the following probability 
\begin{equation}
P = \min \{1, \exp [-\beta (E(\mathbf{X}^\prime) - E(\mathbf{X}))] \}
\end{equation}
where $\beta = (k_\mathrm{B}T)^{-1}$ is the inverse temperature.
\item
Repeat from (ii).
\end{enumerate}
We note that how the trial structure is generated (step ii) has a profound effect on the efficiency of the algorithm. In the original Metropolis paper for rigid discs \cite{Metropolis1953} and in molecular simulations today, the trial step consists of choosing a particle and displacing it by a random amount. This does not work well for solid state systems because they are more or less close-packed, and such trial steps tend to result in a large increase in energy and thus very small acceptance ratios. Instead, our scheme consists of picking two distinct atoms or vacancies and exchanging them. Basically, crystalline systems have well-defined ion sites, so the configuration $\mathbf{X}$ can be represented by a list of site occupations. The Metropolis trial step then consists of picking two sites with different ions, exchanging them to generate the trial configuration $\mathbf{X}^\prime$, translating $\mathbf{X}^\prime$ to DFT input, running the DFT calculation, and extracting the energy from DFT output. The DFT calculation can include local optimization of the structure to account for relaxation effects. In this case, the simulation can be considered to be an implementation of basin-hopping Monte Carlo method \cite{Wales1997}. 

\subsection{Replica exchange}
Even with carefully chosen structure update schemes, the Metropolis algorithm tends to get stuck in local minima, especially when the temperature is low. The RX algorithm, also known as parallel tempering, provides a method to overcome this limitation. The basic idea is to simulate $N_\mathrm{repl}$ copies, or replicas, of the system under study, usually at different temperatures. The simulation on each replica can be performed using MC or MD methods. At certain intervals in the simulations, the temperatures are swapped between replicas (or replicas are swapped between temperatures) based on a Metropolis criterion:
\begin{equation}
P = \min \{1, \exp[(\beta_i - \beta_j) (E(\mathbf{X}_i) - E(\mathbf{X}_j))] \}. \label{eq:RX}
\end{equation}
This allows each replica to travel between low and high temperatures, making it possible to overcome high energy barriers between local minima and provide sampling over a representative set of low-temperature regions in the global configuration space. Since each replica simulation can run independently between exchange attempts, the RX algorithm is well-suited for use on massively parallel cluster supercomputers. Temperature-dependent quantities can be calculated by sampling from the replica with the specified temperature at each MC step.

 \begin{figure}[tb]
\centering
 \includegraphics[width=0.8\columnwidth]{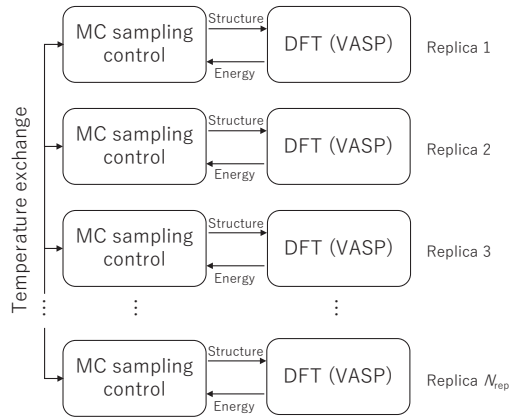}%
 \caption{A schematic of our RXMC-DFT framework in action. $N_\mathrm{rep}$ Monte Carlo processes each spawns density functional calculation processes for obtaining the energy at each Monte Carlo trial step. Temperature exchange is attempted between the Monte Carlo processes at preset intervals.  \label{fig:rxmc_scheme}}
 \end{figure}

\subsection{Parallel implementation}
Our implementation scheme is shown schematically in Fig.~\ref{fig:rxmc_scheme}.
The RXMC part is implemented using Python with mpi4py \cite{Dalcin2005, Dalcin2008, Dalcin2011} for distributed parallel processing using the message passing interface (MPI). The program is started with $N_\mathrm{rep}$ processes, each of which takes care of MC sampling of each replica. At every trial step, each of the $N_\mathrm{rep}$ processes prepares the DFT input files including the trial structure, spawns $N_\mathrm{DFT}$ parallel MPI processes for running a DFT package, collects the results, and accepts or rejects the trial structure based on the Metropolis criteria mentioned above. At preset intervals, temperature exchange is attempted between replicas with adjacent temperatures. More concretely, the first exchange attempt is made between $T_0$ and $T_1$, $T_2$ and $T_3 \cdots$, the second attempt is made between $T_1$ and $T_2$, $T_3$ and $T_4$, and so on where the temperatures $T_0 < T_1 < T_2 < T_3 < \cdots < T_{N_\mathrm{rep}}$. This allows each temperature to travel between all replicas. We note that other exchange schemes including global permutation is possible \cite{Suwa2010,Itoh2013} as long as the balance condition is met; such methods may be implemented in the future. The program code used in this work can be found at \mbox{\url{https://github.com/skasamatsu/py_mc/}}. It is also noted that we use pymatgen \cite{Ong2013} heavily for structure handling and DFT input/output.

\begin{figure*}
 \includegraphics[width=2\columnwidth]{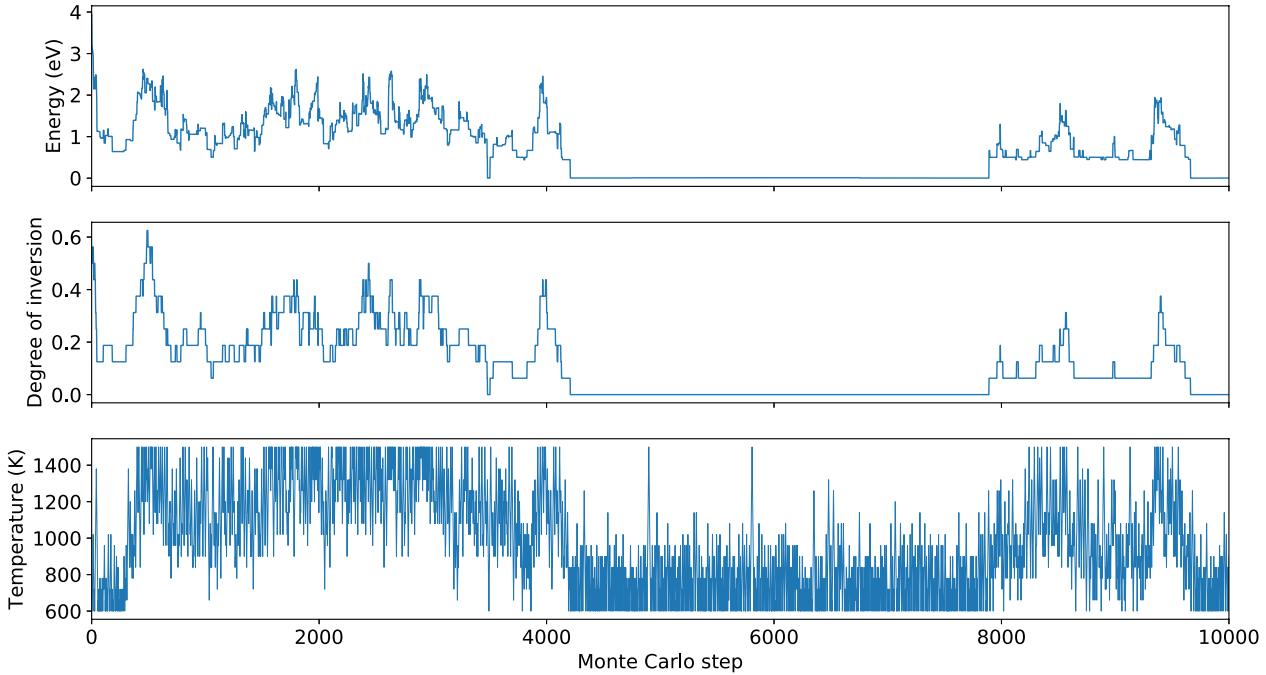}%
 \caption{Total energy vs. the ordered spinel structure (top), degree of inversion (middle), and temperature (bottom) for the first 10,000 Monte Carlo steps in one of the replicas.  \label{fig:rxmc_progress}}
 \end{figure*}

\section{Benchmark calculation on MgAl$_2$O$_4$}
As mentioned in the introduction, we choose to calculate the degree of inversion in the cation sublattice of MgAl$_2$O$_4$ spinel oxide as a benchmark for the feasibility of the present approach. The results can be compared directly with a previous Monte Carlo simulation using a cluster expansion $+$ screened-point charge (CE+SPC) Hamiltonian that was fitted to first-principles results \cite{Seko2014}. If the results turn out to be similar, then we can conclude that it is feasible to bypass fitting models and directly sample from first-principles results to obtain reliable thermodynamical quantities in disordered solids.

\subsection{Computational parameters}
The DFT calculation part is performed using Vienna ab-initio simulation package \cite{Kresse1996}. The calculation supercell contains 16 Mg, 32 Al, and 64 O atoms. The GGA-PBE functional is used to approximate the exchange-correlation energy, and the projector augmented wave method \cite{Blochl1994} is used to describe electron-ion interactions. A plane wave basis set with a cutoff energy of 300 eV is used to express the valence wave functions. We only sample the $\Gamma$ point in the Brillouin zone. Structural relaxation is performed at each Monte Carlo step until the forces are below 0.04 eV/\r{A}.  The above parameters are set not very much for accuracy but rather for speed. We will see if the accuracy is high enough and we can get away with such a setup. It is noted that about 10 ionic steps, which take less than one minute using 2 Intel Xeon Gold 6148 processors, turn out to be enough to optimize the structure using this setup at each Monte Carlo step. This means that we can sample more than 20,000 configurations in one day when using only 16 compute nodes with 2 processors each.

We note that the internal and lattice parameters are simultaneously relaxed in this simulation. The relaxation is performed with a constant basis set that is determined for the initial structure based on the plane wave cutoff energy. This means that if the lattice parameters change significantly in the course of the relaxation, the basis set would correspond to a different cutoff energy for the final structure. To alleviate this problem, we restart the relaxation several times with constant energy cutoff, so that the final structures and energies consistently correspond to the preset cutoff energy of 300 eV.

As for the replica exchange parameters, we use 16 replicas spaced evenly between 600 K and 1500 K. Temperature exchange is attempted at every Monte Carlo step. It is noted that disorder is considered only in the cation sublattice, and the O sublattice stays completely ordered throughout the simulation. The numbers of each of the ion species are also constant throughout the simulation; that is, we are sampling from the canonical ensemble with a total of ${}_{48} C_{16} = 2,254,848,913,647$ possible configurations (we do not consider any symmetry for reducing the number of independent configurations). We compared two simulations starting from randomly initialized configurations and ordered spinel configurations to examine the possibility of sampling bias due to the choice of the initial configurations. The randomly initialized simulation was carried out for 16,750 steps, while the simulation starting from the ordered structure was carried out for 13,000 steps.

\begin{figure}
 \includegraphics[width=\columnwidth]{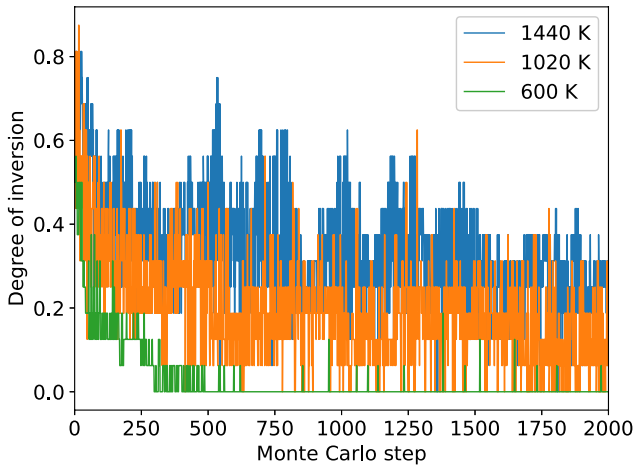}%
 \caption{The degree of inversion of the replica at specified temperatures (green: 600 K, orange: 1000 K, blue: 1440 K) for the first 2,000 Monte Carlo steps.  \label{fig:T-DOI}}
 \end{figure}

\begin{figure}
 \includegraphics[width=\columnwidth]{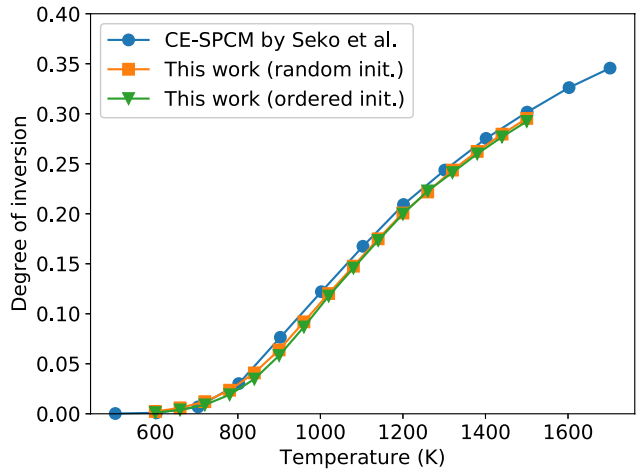}%
 \caption{Comparison of the average degree of inversion as a function of temperature calculated by CE-SPCM (Seko et al. \cite{Seko2014}; circles) and the present approach (randomly initialized: squares; initialized with an ordered spinel structure: triangles).   \label{fig:DOI}}
 \end{figure}

\subsection{Results}
To assess the efficiency of the algorithm, we first examine the temperature, energy, and degree of inversion (DOI) histories of one of the replicas in the randomly initialized simulation (Fig.~\ref{fig:rxmc_progress}). The DOI is defined as the ratio of Mg sites that are occupied by Al ions. It is noted that the DOI is 0 for the perfectly ordered spinel structure and $2/3$ when the cation sublattice is completely disordered. Perhaps trivially, low energy corresponds to low DOI and low temperature. The lowest energy corresponds to the ordered spinel structure with DOI of zero, and we find that when a replica finds this structure, it tends to stay there for a long time, or in other words, get stuck. This is because of an ``energy gap'' of about 0.4 eV vs.~the second lowest energy structure, which can only rarely be overcome. Temperature exchange according to Eq.~\ref{eq:RX} does not help much, since the replica with the lower energy is unlikely to travel to higher temperatures. This results in a rather acute degradation of the sampling efficiency, and is a known deficiency of RXMC method for first-order phase transitions \cite{Machta2011}. Multicanonical techniques \cite{Berg1992, Wang2001} are known to perform better in this regard. The application of such alternate sampling techniques for direct first-principles sampling is an important direction for future works.

Nevertheless, having 16 replicas seems to alleviate this problem somewhat. We plot the degree of inversion of the replica with temperatures of 600 K, 1020 K, and 1440 K at each Monte Carlo step in Fig.~\ref{fig:T-DOI}. We find that the 600 K series finds the lowest-energy ordered spinel structure within 300 steps, showing that the method is working rather well not only as a sampler but also as an optimizer considering the fact that it found the single most stable structure out of 2,254,848,913,647 possible configurations. The higher-temperature series seem to be well equilibrated after $\sim 1500$ steps. 

To be on the safer side, we threw out the first 3000 steps and performed averages over the remaining steps to obtain the temperature-dependent expectation value of the DOI as shown in Fig~\ref{fig:DOI}. Calculated DOI from the two simulations starting from randomly initialized configurations and that starting from the ordered spinel structure coincide within the size of the symbols in the figure. This suggests that enough MC steps were performed to obtain results that do not spuriously depend on the initial configurations. Most importantly, our results are virtually identical to the results from the CE+SPC model, strongly suggesting that indeed, we can perform sufficient sampling directly from first principles. The slight but systematic difference (our results predict a slightly smaller DOI) may be due to the smaller unit cell in our calculation, or due to the less accurate DFT calculation parameters that were chosen in favor of calculation speed.

\section{Conclusion}
In this work, we presented a scheme for performing thermodynamical sampling of disorder in materials by combining the replica exchange Monte Carlo method directly with structural relaxation and energy calculation using DFT. The multilevel parallelism in the scheme allows for efficient execution on today's massively-parallel supercomputers. We successfully calculated the temperature-dependent disorder in spinel oxide using this scheme and demonstrated that such direct sampling of DFT results is indeed feasible. Hence, the present scheme provides a way to sample materials disorder in systems where reliable model fitting has been very difficult, if not impossible. This admittedly brute-force approach will allow us to tackle problems in technologically relevant systems such as disordered oxides and electrochemical interfaces that were previously impossible to handle due to the number of first-principles calculations necessary for sufficient sampling.

\section*{References}
\bibliographystyle{iopart-num}
\bibliography{article}

\end{document}